# STARTUPS AND STANFORD UNIVERSITY

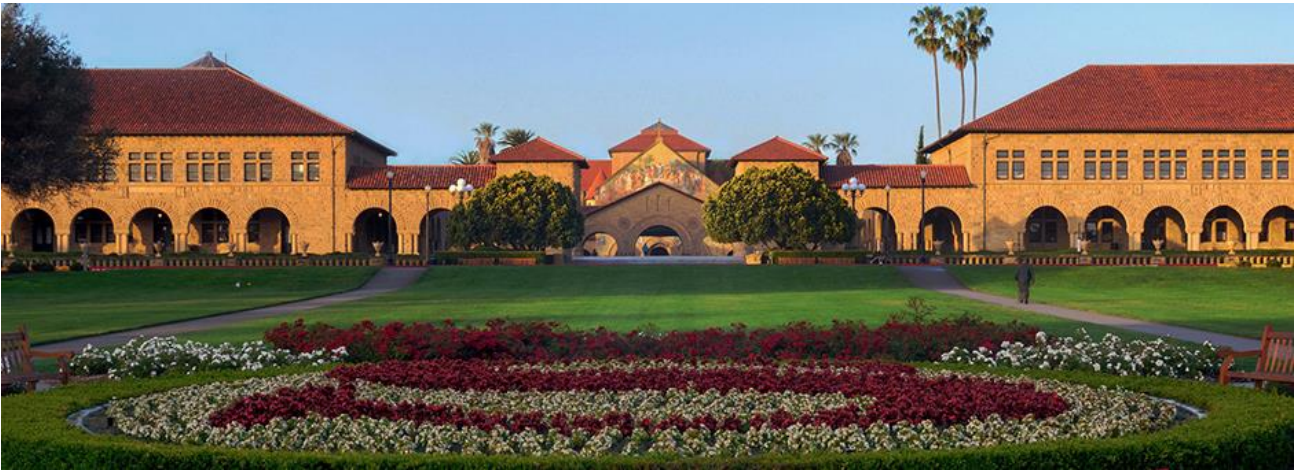

Stanford University is one of the best universities in the world. Its beautiful campus in the middle of Silicon Valley welcomes brilliant students in all fields of sciences and humanities as well as the best professors and researchers. Nearly as well-known, the university has been at the origin of some of the most famous startup success stories such as Hewlett Packard, Sun Microsystems, Cisco, Yahoo, Google, VMware, Instagram or YouTube, just to name a few. Entrepreneurship is however much more than story-telling and indeed Silicon Valley has been a huge terrain for academic research in economics, entrepreneurship and innovation. Stanford University may have been less so. This report analyzes more than 5'000 companies and also more than 5'000 founders with the ambition to give a renewed point of view on this unique creation of value.

## Entrepreneurship, Startups and Spinoffs

Entrepreneurship and Innovation have probably become an important topic of research with seminal work of Joseph Schumpeter, the "Prophet of Innovation" [1] and his concept of Creative Destruction. His huge research corpus explored the surprising importance of small, but fast-growing firms in economics. Not all companies are startups or spinoffs. Indeed the definition of a startup is still not clear. According to Wikipedia, a startup company (startup or start-up) is an entrepreneurial venture which is typically a newly emerged, fast-growing business that aims to meet a marketplace need by developing or offering an innovative product, process or service. Although this can be seen as a good definition, Steve Blank, a Silicon Valley serial entrepreneur, has come with a more recent and probably better definition:

*Startups are temporary organizations designed to search*
*for a scalable and repeatable business model.*

In complement, a University spinoff is a company founded by members of the university. Whether a spinoff is a startup or not depends upon its specific features. One can refer to Academic Entrepreneurship, one of the classical references about academic spinoffs [2].

## Stanford University

Stanford University was founded in 1899. It would certainly be more artificial to give a birthdate for the startup phenomenon. Silicon Valley faces a similar challenge. Whereas 1957 is commonly accepted for the premier technology cluster, some experts claim that 1939 for the foundation of Hewlett Packard or even 1909 for the creation of Federal Telegraph in Palo Alto would be better foundation years. There is no doubt however that 1957 with the beginning of the space exploration, the development of the Cold War and the foundation of Fairchild Semiconductor, maybe the first startup ever, has been a critical year for technology innovation. In her remarkable book [3], Rebecca S. Lowen shows how Stanford was transformed thanks to the federal funding for science after Second World War without forgetting the central figure of Frederick E. Terman. The fact that Stanford is in the middle of Silicon Valley was certainly a strong reason for that transformation and success, but the argument could be reverted to explain the success of Silicon Valley thanks to Stanford, a kind of chicken and egg situation. It is worth mentioning though that the relationships between Stanford and Silicon Valley were complex and cannot be described by simple two-way flows [4].

## Academic Startups and Spinoffs

In the decades following the 50s and 60s, startups and academic spinoffs have become an extraordinary phenomenon. A great even if not well-known analysis of Silicon Valley startups [5] shows that the region was home to more than 22'000 high-tech firms in 2003 and more than 29'000 such firms had been created during the 90s (with a sharp decline thereafter). Most universities have published some analysis on their startups, for example at MIT [6], at Stanford [7] or in Switzerland at ETH Zurich [8], [9] and EPF Lausanne [10]. In his analysis [7], Eesley claims that "39'900 active companies can trace their roots to Stanford. If these companies collectively formed an independent nation, its estimated economy would be the world's 10th largest. Extrapolating from survey results, those companies have created an estimated 5.4 million jobs and generate annual world revenues of $2.7 trillion."

This report analyzes the performance of more than 5'000 firms which have a link to Stanford University. For more information, go to section "About the Data" at the end of the report. Of course entrepreneurship is not only about technology companies, but in Silicon Valley, and in particular at Stanford, most companies are high-tech as shows figure 1. Also many firms are service companies with no product offering. About 30% of the firms studied here are in that situation (see Appendix for more graphics). Overall high-tech firms related to information technologies represent more than 50% of the sample. They include firms selling hardware (HW) products such as semiconductors, computers, telecom equipment and electronics as well as software (SW) including multimedia and Internet technologies. It must be mentioned here that Internet services are considered as part of these software firms (showing the difficulty in classifying firms by domain of activities)

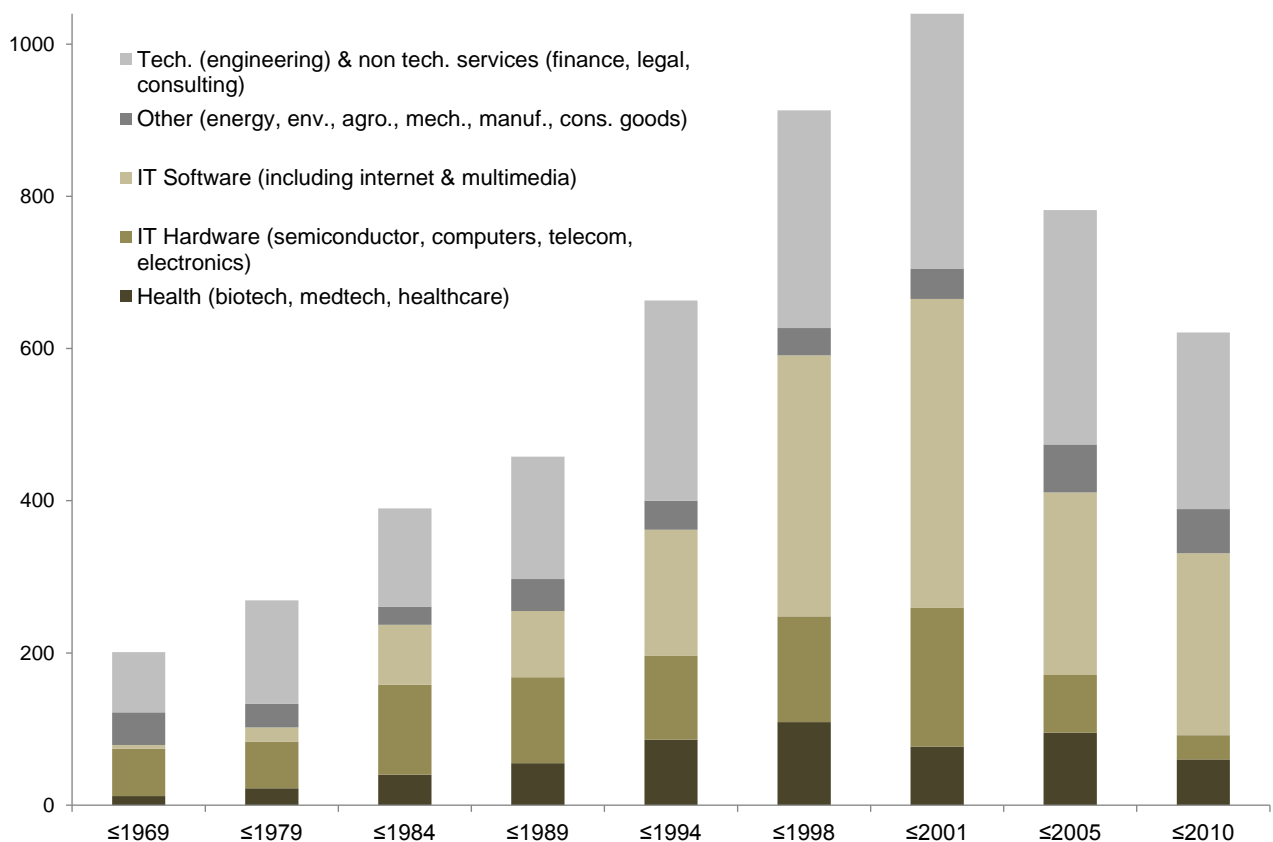

*Figure 1: The Stanford startups by period of foundation and domains of activity*

## Status of Firms

Firms are not eternal and indeed their life expectancy is quite short. Zhang [5] shows that about half of both service and non-service firms had died 10 years after their creation. About a third of the firms had stopped their activities and surprisingly the ratio increases over time. The simplest explanations are either a bias in the database for early years or an increase in failure with the entrepreneurship fever which accompanied the Internet development. A quarter had been acquired (M&A) and a non-negligible part had gone public before at some point (6% in total). Another third was still private whereas a tiny 3% were publicly quoted.

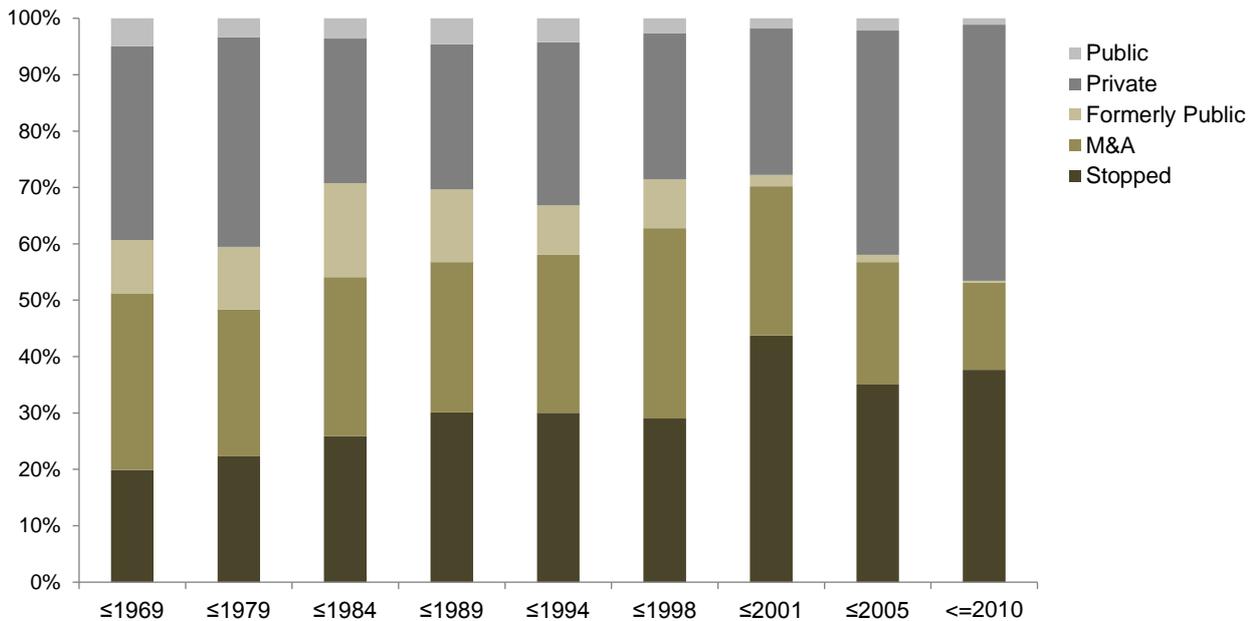

*Figure 2: Status of firms with period of foundation*

So what is the life expectancy of these firms as private companies? Figure 3 shows the results. An overall average of 6.9 years before a cessation of activity, 7.8 years before being acquired and 7.3 years before going public. (For public companies, the time span represents years from foundation to IPO). These averages hide however a regular decrease until 1998 with more stable values thereafter. Table i in Appendix adds more information with a more granular analysis by fields.

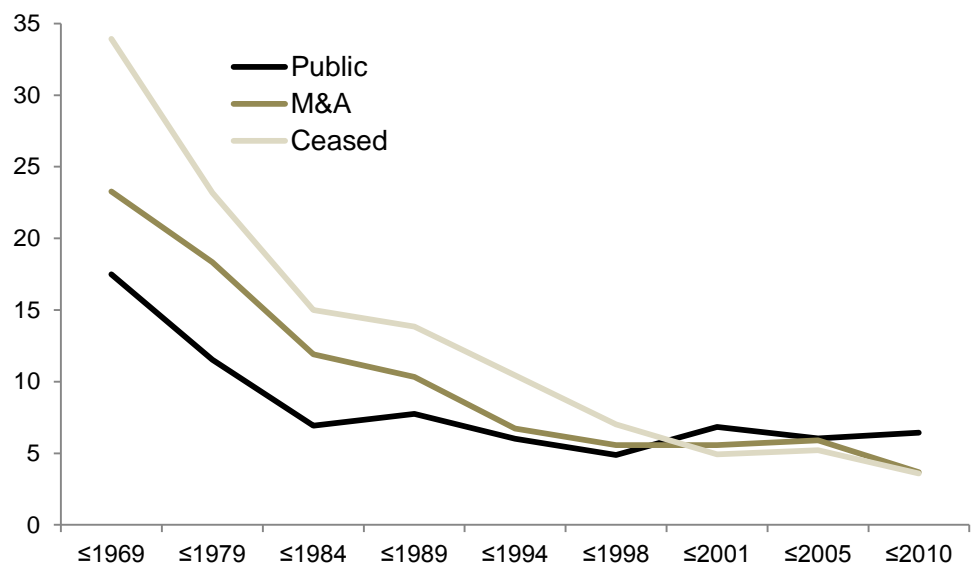

*Figure 3: Average time (in years) before exit*

# Value Creation

Value creation is a difficult analysis to make for private companies. Most of these companies do not communicate about their numbers when they still exist and very little is known when they disappear. Public companies are much easier to analyze thanks to the documents they publish on a regular basis from their initial public offering (IPO) onwards. In-between some "relative" value creation is known when such companies are acquired with a disclosed value. A systematic analysis was done for public companies as well as for companies which had gone public at some point. The M&A transaction values were also compiled when publicly available.

## Public Companies

There were 148 public firms as of July 2017. The following table describes some of their features.

| Field of Activity | # of firms | Revenues 2016 ($B) | Income 2016 ($B) | Employees | Market Cap. July 2017 ($B) |
|---|---|---|---|---|---|
| Health (biotech, medtech, healthcare) | 42 | 20.7 | 1.0 | 53'000 | 174 |
| IT Hardware (semiconductor, computers, telecom, electronics) | 50 | 263.7 | 33.0 | 757'000 | 662 |
| IT Software (including internet & multimedia) | 34 | 229.9 | 42.8 | 253'000 | 1'126 |
| Other (energy, env., agro., mechanical, manuf, cons. goods) | 11 | 61.2 | 4.9 | 263'000 | 119 |
| Tech. (engineering) & non tech. services (finance, legal, consulting) | 11 | 56.4 | 7.1 | 113'000 | 122 |
| Total | 148 | 632.2 | 89.0 | 1'440'000 | 2'205 |

Table 1: Value creation by public companies

## Former Public Companies

There had been many more firms going public. In addition to the 148 existing public firms, another 333 had gone before being acquired (279), before stopping their activities (36) or becoming private again (18). The next table compiles the average value at the IPO and 12 months after the IPO.

| Fields | # of firms | Value at IPO ($M) | Value 12 months after IPO ($M) |
|---|---|---|---|
| Health | 97 | 202 | 179 |
| IT Hardware | 111 | 619 | 379 |
| IT Software | 102 | 906 | 1'048 |
| Other | 11 | 248 | 347 |
| Tech. & non tech. services | 12 | 333 | 393 |
| Overall | 333 | 563 | 521 |

Table 2: Average market capitalization of companies which went public

Values at IPO are not sufficient to describe the value creation and even if the value after 12 months is also a limited snapshot, it has the advantage of giving usually a more accurate picture of the real value creation.

## Acquired Firms (M&A)

Most startups do not go public. Again about 3% are public and another 6% had been public at some point. Some stay private but many are also acquired. As of 2017, 1'419 firms (25% of the total) had been acquired. The known value of these acquisitions reaches the total amount of $92B. The acquisition value is known for 533 firms only which gives an average of $173M.

| Fields | # of firms | M&A Value ($M) | Total M&A Value ($B) |
|---|---|---|---|
| Health | 85 | 142 | 12 |
| IT Hardware | 195 | 217 | 42 |
| IT Software | 218 | 148 | 33 |
| Other | 12 | 127 | 1.5 |
| Tech. & non tech. services | 23 | 162 | 3.7 |
| Overall | 533 | 173 | 92 |

*Table 3: Average and cumulative M&A transactions*

Again Appendix further describes the data by field but also gives additional information about M&A transactions for companies which had gone public (see tables v to viii).

## Venture Capital

Although not created in Silicon Valley, Venture Capital (VC) is co-substantial to the San Francisco Bay Area. Numerous books and articles (including academic ones) describe this very unique investment activity and one of the most accessible one is documentary film SomethingVentured [11]. However venture capital remains controversial in many places, even in Silicon Valley. Is it a necessary component of innovation and high-tech entrepreneurship? Is the value added just money or more? This is not the place to analyze what venture capital brings. Again here are some facts and figures.

In total, 1'676 firms had raised money from investors. This was mostly done with venture capital (1'614 used VC and only 62 firms did not have a VC identified). This is 24% of the full sample and more interestingly 1'597 startups out of the 3'103 of the health and information technologies fields (51%). In total, this means $12B for the Health sector, $22B for IT Hardware, $27B for IT Software and a smaller $3.6B for the other fields.

| Fields | # of firms | Average Amount Raised ($M) | Total Amount Raised ($B) |
|---|---|---|---|
| Health | 342 | 40 | 12.5 |
| IT Hardware | 503 | 47 | 22.1 |
| IT Software | 752 | 39 | 27.7 |
| Other | 50 | 73 | 3.2 |
| Tech. & non tech. services | 29 | 19 | 0.4 |
| Overall | 1'676 | 43 | 65.9 |

*Table 4: Amounts of investments by fields*

Can we bring some additional food for thought to the impact of venture capital? It is possible to compare the value creation by companies which raised venture capital and those which did not. For the public companies 105 startups are identified with VC and 41 are not. In terms of job creation it is about 750'000 jobs for the ones which did not have VC. Hewlett Packard (244'000), Flextronics (200'000) and The Gap (135'000) are the main job creators for these. In comparison, 660'000 are currently employed by public companies which received venture capital. Google (74'000), Cisco (73'000) and Baidu (46'000) are among the most famous. The VC-backed companies generate however more revenues, profits and value for shareholders, but less than pro-rata their number.

| Type of firms | # of firms | Revenues 2016 ($B) | Income 2016 ($B) | Employees | Market Cap. July 2017 ($B) |
|---|---|---|---|---|---|
| VC- backed | 105 | 356 | 57 | 678'000 | 1'641 |
| Non-VC backed | 43 | 274 | 31 | 750'000 | 559 |

*Table 5: Public companies and venture capital*

The same analysis can be done for formerly public companies as well as for M&A transactions.

| Type of firms | # of firms with an IPO | Market Caps after 12 m. ($B) | Average Value ($M) | # of M&A Transactions | M&A Value ($B) |
|---|---|---|---|---|---|
| VC- backed | 267 | 157 | 589 | 717 | 70 |
| Non-VC backed | 66 | 15 | 243 | 700 | 21 |

*Table 6: Past public companies, M&A transactions and venture capital*

A different illustration is given by the following figure. Here the value creation sums the IPO and the M&A values of VC-backed companies, compared to the amounts raised by companies founded during the same period (these are possibly different companies). The M&A transactions include only companies which never went public and the public values are taken at IPOs to avoid possible double counting. The ratios are above 7x before 1998 and below 2x since 2001. This seems to indicate an evolution in value creation in recent years, despite an overall huge success for venture capital.

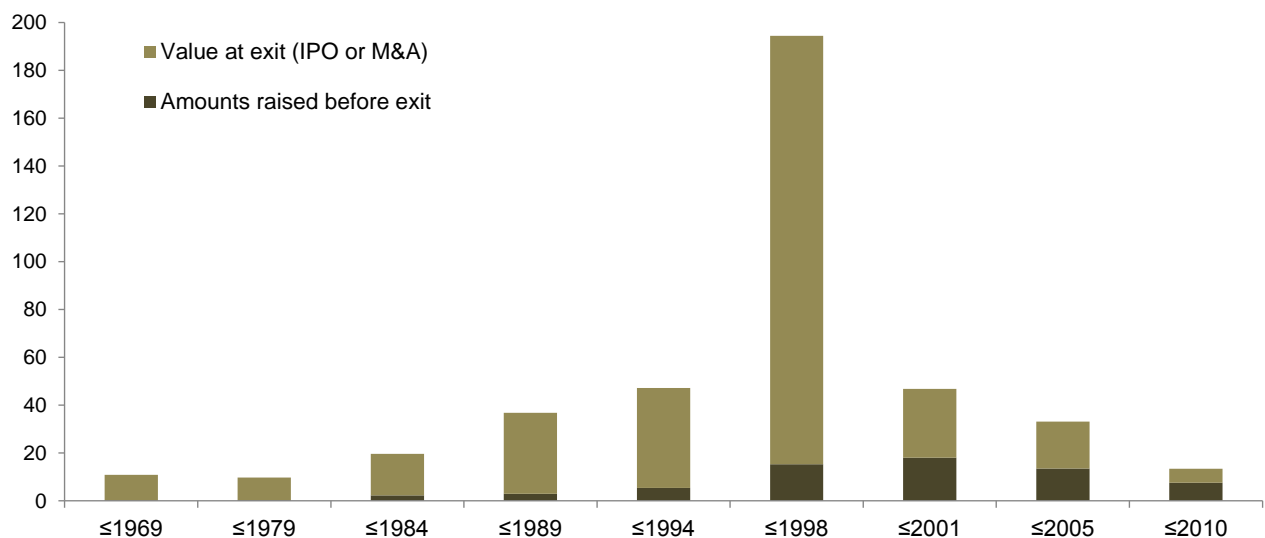

*Figure 4: Value creation at exit vs. amounts raised (in $B)*

Another interesting feature is the life expectancy of firms vs. venture capital as expressed next:

| Type of firms | Overall | Public | Formerly public | M&A | Stopped |
|---|---|---|---|---|---|
| VC- backed | 6.3 | 7.7 | 5.9 | 6.4 | 5.9 |
| Non-VC backed | 8.2 | 13 | 8.9 | 9.5 | 7.3 |

*Table 7: Life expectancy of firms (in years) and venture capital*

## Active Venture Capitalists around Stanford

Startups and Stanford UniversityThe database includes about 700 firms in the finance field, and about 200 venture capital firms. These include Alta Partners, Asset Management, Benchmark Capital, DFJ, Index Ventures, Khosla Ventures, Mayfield, MPAE, Sutter Hill, USVP. More importantly, more than 6'000 VC firms are mentioned in the 1'614 invested companies. The most active firms are given in the next table.

| VC Firm | # Inv. | VC Firm | # Inv. |
|---|---|---|---|
| Kleiner Perkins Caufield & Byers | 141 | Sequoia Capital | 125 |
| New Entreprise Associates | 114 | Mayfield Fund | 93 |
| Draper Fisher Jurvetson | 79 | Institutional Venture Partners | 65 |
| Accel Partners | 65 | U.S. Venture Partners | 56 |
| Mohr Davidow Ventures | 56 | Menlo Ventures | 54 |
| Sutter Hill Ventures | 53 | Venrock Associates | 50 |
| InterWest Partners | 48 | Greylock Partners | 48 |
| Benchmark Capital | 47 | Morgenthaler Ventures | 39 |
| Norwest Venture Partners | 37 | Bessemer Venture Partners | 36 |
| Oak Investment Partners | 35 | Alta Partners | 33 |
| Hambrecht & Quist | 31 | August Capital | 31 |

*Table 8: Most active VC firms*

This is a well-known fact: the density of the VC industry in Silicon Valley is an important networking element. Entrepreneurs, investors and managers are closely connected which makes Silicon Valley a unique entrepreneurial ecosystem.

## Geography of Startups

The vast majority of firms is or was based in California as table 8 shows. Even if Silicon Valley was not specifically studied, it can be added that the majority is based around San Francisco. The rest of the USA adds another 1'438 firms. Eastern Asia counts 134 companies and Europe 123. 442 companies were not located (not all state corporation registries are open access) and another 143 are incorporated in Delaware, which does not mean a physical location in that state.

| Geography | # Firms | Geography | # Firms |
|---|---|---|---|
| California | 3'424 | South & Central America | 50 |
| New York | 171 | Canada | 17 |
| Massachusetts | 152 | China | 37 |
| Washington | 129 | Taiwan | 22 |
| Texas | 105 | Hong Kong | 22 |
| Colorado | 75 | Japan | 17 |
| Illinois | 67 | Korea | 9 |
| Oregon | 56 | Other Asia & Oceania | 27 |
| Florida | 35 | Israel | 12 |
| Pennsylvania | 30 | Middle East & Africa | 18 |
| Arizona | 29 | United Kingdom | 41 |
| Other East Coast | 204 | France | 24 |
| Delaware | 143 | Germany | 10 |
| Other US States | 242 | Switzerland | 10 |
| Unknown | 442 | Other Europe | 38 |

*Table 9: Geography of startups*

## Spinoff or not Spinoff?

In a startup guide published in 2012, the Stanford Office of Technology Licensing (OTL) explained that "with all of this entrepreneurial activity, some people are surprised to learn that only about 8-12

July 2017                                                                                                           Page 7

of OTL's licenses per year (approximately 10% of its total licenses) are to start-up companies." Indeed, our database counts 222 spinoffs only, founded between 1965 and 2010.

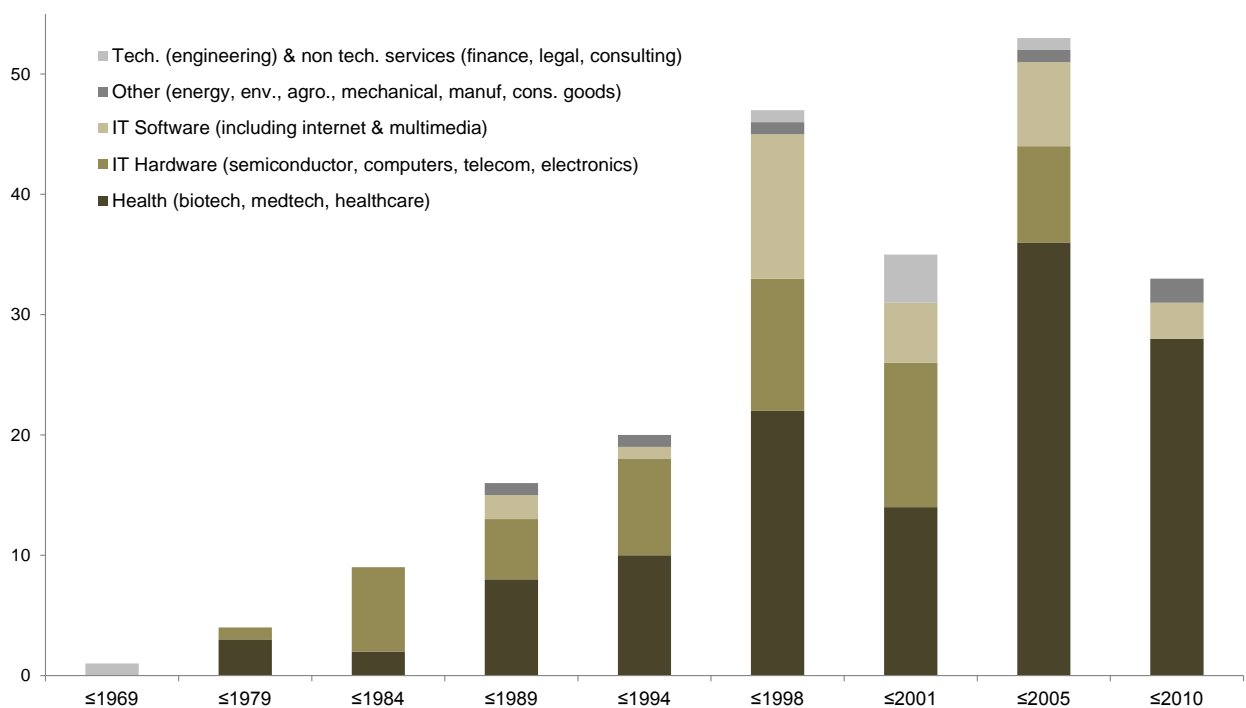

Figure 5: Stanford spinoffs by period of foundation and fields

Stanford OTL has a valid definition of spinoff, but it is possible to consider a broader definition. If a spinoff is an entity created from an institution, formal intellectual property (IP) is only one of the possible sources of creation. People creating a company during their activity at Stanford usually benefit from that environment even if they do not create formal IP. A famous example is the Google vs. Yahoo situation [16]: *"The Google and Yahoo! stories illustrate the application of Stanford's Patent and Copyright Policies to real-life examples. Jerry Yang and David Filo disclosed their software to Stanford, requesting that Stanford confirm that Stanford did not have an ownership interest in the technology. Yang and Filo were Ph.D. students at Stanford and had used Stanford computers (which is usually considered to be incidental use) to develop the software; their professors confirmed that their invention was not related to their university responsibilities as students. Based on this information, Stanford did not claim ownership to what became the Yahoo! search engine. In contrast, Sergey Brin and Larry Page had worked on a search engine for many years. Because the students had been paid by a government contract in the course of their research to satisfy their Ph.D. degree requirements, under both Stanford's Patent and Copyright policies Stanford had ownership to the software, that is, the written code. In addition, Stanford filed a patent on the method of ranking Web pages in order to improve searches. After trying to find the best licensee, Stanford determined that these inventors were in the best position to develop the invention effectively, and so Stanford licensed the technology to their company, Google."* There were other cases where the status as a spinoff was a source of heated debate. Again, this is not the place to develop the topic.

## Founders

"Founders" does not have a strict definition and it even happens that some individuals claim to be founders of firms that other founders would not agree with. This being said, it should also be added that the data gathered here mostly include founders with a Stanford affiliation (see the section About the Data). However 55 Stanford spinoffs (out of the 222) do not even have any Stanford founders identified as the licensed intellectual property seems to be the only link with the university. Our database counts 5'181 unique individuals identified as founders.

## Number of Stanford Founders per Firm

As an introduction to the founders' analysis, figure 6 shows the percentage of companies relatively to the number of founders. The reader should be cautioned again. The figure does not say that about or more than 80% of the companies only have one founder in all categories. It says that 85% of the companies only have one *Stanford* founder. A further analysis is shown in figure 7 and compares the number of founders with the amount of money raised, the value at IPO and M&A transactions as well as for public firms in 2016 their market capitalizations and employment. The data seems to confirm a fact that figure 6 did not show, that is more founders help in value creation. This seems to be particularly reinforced for the long term, i.e. existing public firms in July 2017.

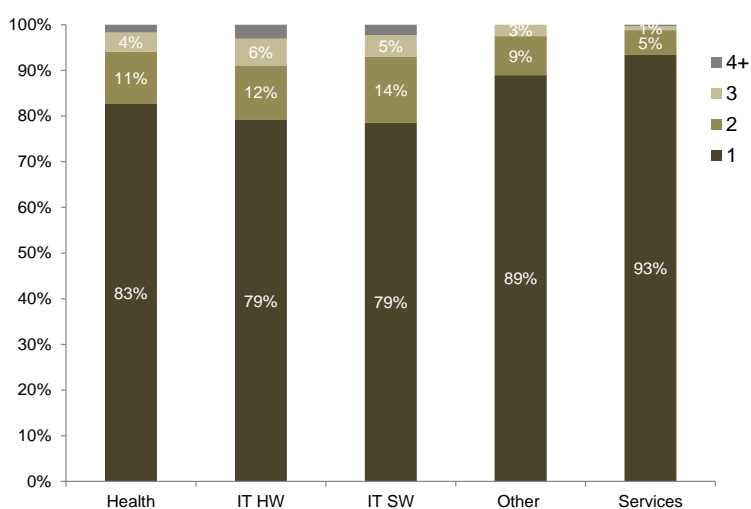

*Figure 6: Ratio of number of Stanford founders in firms*

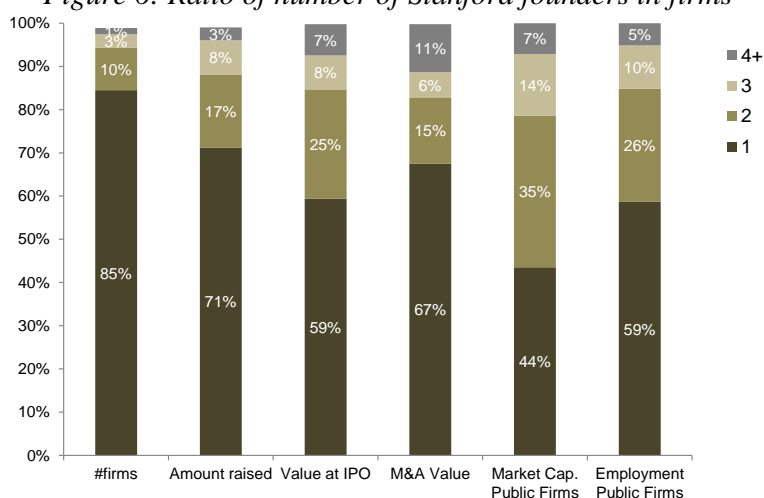

*Figure 7: Ratio of number of Stanford founders in firms and value creation*

## Academic Background of Founders at Stanford

The Stanford background is illustrated by the next figure on founders. Again, this is limited by the fact that these are the Stanford affiliations only. A founder might have a PhD, MS, MBA or even professor position from another university. Figure 8 shows a close to equal balance between MBAs and Masters of Sciences (MS) and a smaller number of PhDs or professors except in the "deep" technology fields (HW or health).

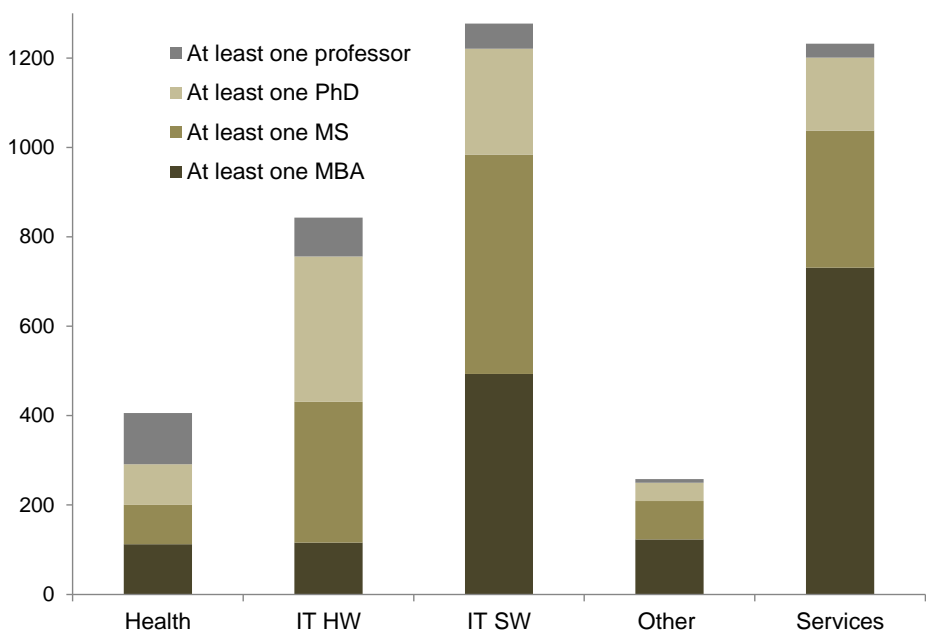

*Figure 8: Background of founders*

This section would deserve much more analysis, in particular because of the complexity of the associations of founders. However a simple description analysis of the value creation follows.

|  | #firms | Amount raised | Value at IPO | M&A Value | Market cap. of public firms | Employment of public firms |
|---|---|---|---|---|---|---|
| At least one professor | 297 | 7.4 | 34.8 | 74.3 | 221 | 115'000 |
| At least one PhD | 857 | 16.4 | 67.5 | 107.5 | 288 | 148'000 |
| At least one MBA | 1'575 | 16.8 | 60.1 | 94.4 | 358 | 445'000 |
| At least one MS | 1'264 | 17.3 | 148 | 145 | 1'269 | 370'000 |

*Table 10: Value creation ($B) and background of Stanford founders*

## Serial Entrepreneurs

The topic of serial entrepreneurs would probably require a dedicated study and the interested reader may want to read an earlier analysis from a subset of this database [14]. With 1'071 serial entrepreneurs, our database contains 80% one-time entrepreneurs and 20% multiple founders.

| # firms by serial founder | # serial founders | # firms by serial founder | # serial founders |
|---|---|---|---|
| 1 | 4'110 | 5 | 27 |
| 2 | 731 | 6 | 17 |
| 3 | 214 | 7+ | 7 |
| 4 | 75 | | |

*Table 11: Serial entrepreneurs*

We will only mention some data about value creation from these two types of founders. The table gives the amount of money raised, the M&A and IPO values for one-time founders (Serial Index = 0) and for serial entrepreneurs with their 1st to 4th venture as well as above the 2nd one.

| Serial index | Amount raised | M&A Value | Value at IPO | Value 12 m. after IPO |
|---|---|---|---|---|
| 0 | 42 | 481 | 690 | 826 |
| 1 | 28 | 739 | 489 | 481 |
| 2 | 47 | 692 | 848 | 681 |
| 3 | 52 | 412 | 894 | 1'601 |
| 4 | 58 | 232 | 814 | 786 |
| 2+ | 51 | 591 | 848 | 809 |

*Table 12: Average value creation ($M) and serial entrepreneurs*

The table does not probably show enough and even if Appendix adds information, more research would be required. There might for example be a kind of trust effect in favor of successful serial founders that might create a bias in both investments and perceived value creation.

## Entrepreneurship and Academic Life

A major question is the real impact of Stanford in that entrepreneurial activity. The report has already touched the topic through the spinoff definition. Another major element might have been addressed earlier in this report, i.e. the timespan between the academic position and the entrepreneurial activity. Figure 9 gives the number of firms founded vs. the number of years between the activity at Stanford and the foundation of the venture. The reason why the number is high for year 0 comes from the fact that professors and other Stanford employees contribute in a unique manner to that specific year.

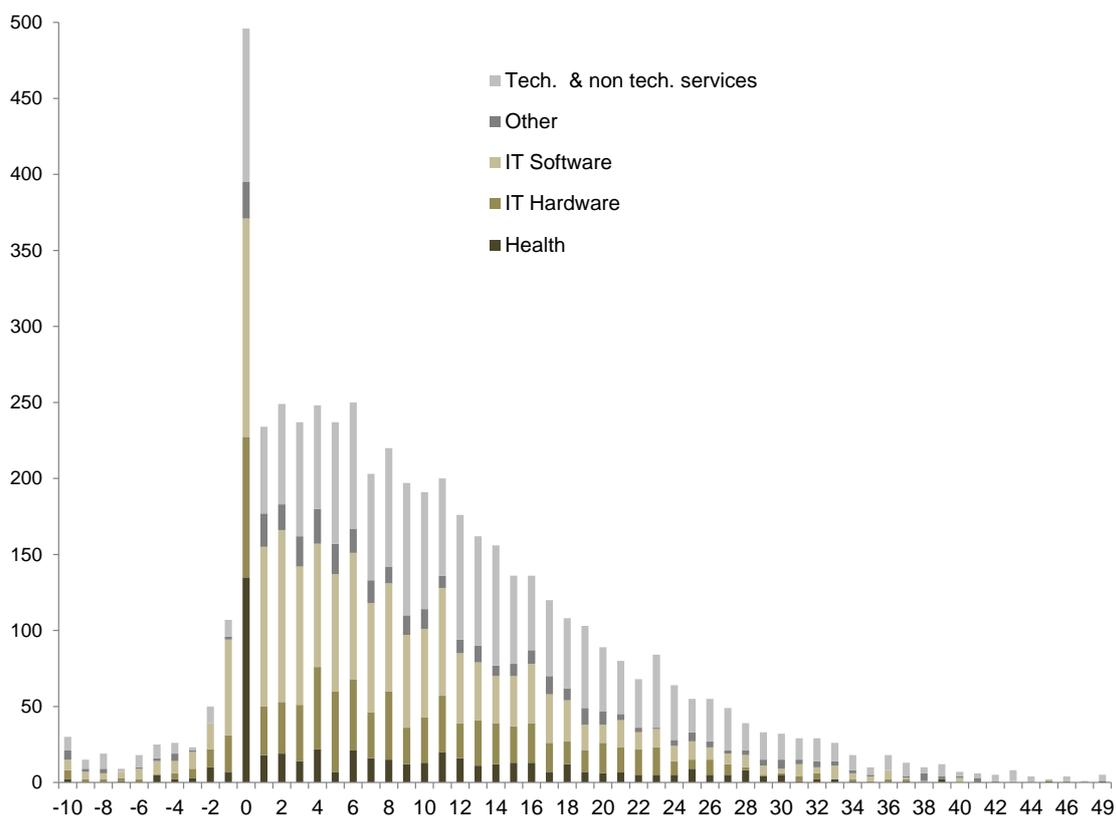

*Figure 9: Years between academia and entrepreneurship*

It remains difficult to say what the Stanford impact is. Certainly the direct impact decreases with years, with the possible remaining influence of the unique experience former students gained during their stay. Even in the first or second year after leaving, the influence remains high but then probably decreases sharply. Still, it is interesting to have a look at the value creation relatively to these years. The years have been grouped to give a similar number of firms per period, i.e. Y<0, Y0, then groups of increasing year spans. We illustrate this point with the amounts raised, M&A values as well as values at IPO, and finally current public companies market capitalizations and employment. It is quite interesting to notice that there is a clear value creation by firms founded during academic activity compared to future years.

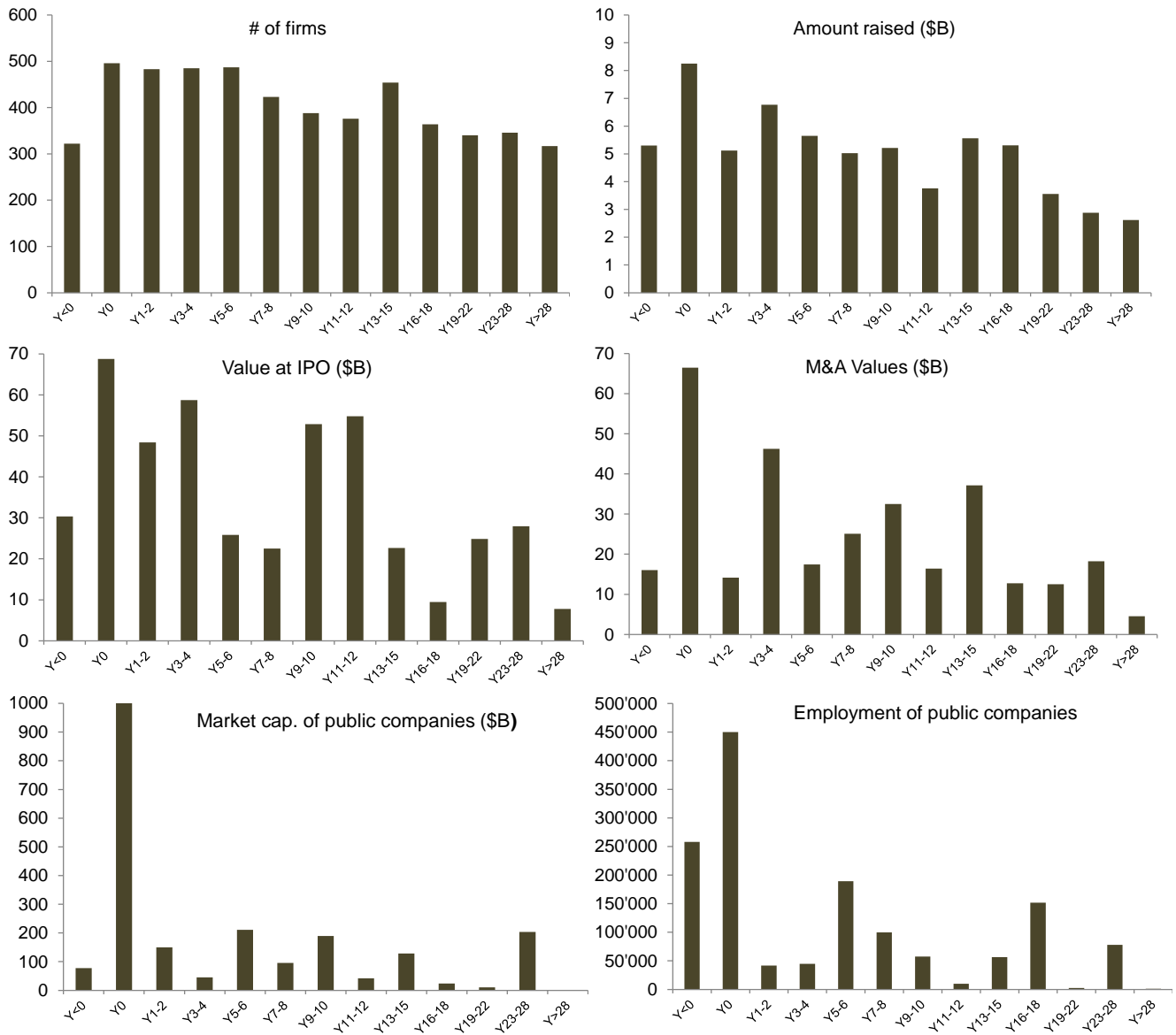

*Figure 10: Value creation vs. years from academia to foundation*

# Conclusion

Let us begin this short conclusion with what this report is not. This is not a traditional research report and the information provided should be considered as a work in progress with a lot of possible research directions. Although Silicon Valley and Stanford are famous for many success stories, this report is not about specific individual cases. There are also important topics which are not addressed, such as the role of minorities, gender and migrants in entrepreneurship.

The report does not analyze the Stanford ecosystem and how so much value creation was made possible. Technology clusters have been a much researched topic but there is still no recipe about how such successful ecosystems can be built. Even worse, it is not clear whether the value creation around Stanford is not just a combination of human (entrepreneurs', inventors', managers', investors' expertise and talent) and financial (research funding, corporate funding, venture capital) resources which slowly built an ideal and optimized culture targeting high-tech innovation, with a secondary-only role of institutions and support mechanisms.

There has always been a debate about how exceptional Silicon Valley and in particular Stanford was. Indeed, this huge value creation is mainly created by a small number of high-flyers and the failure rate remains high even in this highly successful region. The report still shows a higher than usual rate of success and a very high entrepreneurial activity in high-tech innovation. In 60 years, innovation accelerated and resources available increased decade after decade only slowed down slightly during regular crises such as the internet burst in the early 2000s and despite many predictions of the contrary. The region has always looked saturated in many dimensions and the flow of innovation has seemed to slow down recently, except in less technology-oriented ventures such as mobile and internet consumer services. How will it develop in the future is obviously impossible to predict. Human and financial resources will not disappear any time soon and the region stays a powerful magnet. Therefore a revisited analysis of the situation in a decade or so should be very interesting.

## About the Data

The main source of raw data was the Wellspring of Innovation (web.stanford.edu/group/wellspring/), a web site last updated in 2011. The web site gives a list of more than 5'000 companies with their Stanford-affiliated founders. The only additional information it gives was the web sites of the corporations when available. The author also received in 2008 information on Stanford spin-offs and related companies from the Stanford Office of Technology Licensing. The raw output is a list of 5'658 companies with (after some analysis) 5'181 founders. Most of these companies were therefore founded before 2010, which is a sufficiently interesting element as the life expectancy of the sampled firms is about 7 years. The reader should be aware that the link between Stanford and these founders is very diverse. Some created their spin-off while at Stanford using intellectual property created during their professional activity in a Stanford laboratory, while others might have been students many years before creating their firm, in a field which might have no link with their Stanford diploma.

All additional data was obtained by the author from a variety of sources, mostly public ones. In addition to individual web pages of companies and founders, these sources include the sites for corporate entity search of the Secretary of States (www.secstates.com) as well as foreign registers - more rarely though -, the Link Silicon Valley (www.linksv.com) dedicated to Silicon Valley companies, founders, investors and their relative connections, the Internet Archive (www.archive.org), the Securities and Exchange Commission (www.sec.gov) as well as the Wharton Research Data Services (www.whartonwrds.com) for public companies, Crunchbase (www.crunchbase.com) for private companies. For the founders the main source of information was LinkedIn (www.linkedin.com) and the Stanford Alumni database (alumni.stanford.edu).

The work began in 2009 after the author wrote a first book about Silicon Valley [12]. That book contains a dedicated chapter to the spinoffs created at Stanford ISL – the Information Systems Laboratory. This initial work was followed by academic papers about Stanford startups [13], Serial entrepreneurs [14] and a slightly related study about the Age of founders [15]. The interested reader will find more information about all these sources in article [13]. The raw analysis was completed in July 2017, which represents the date of the status of all companies. The author must warn the reader that an analysis of such scale done by a single individual is subject to mistakes and inaccuracies. The author hopes that these possible mistakes are made unimportant with the amount of data collected. This remains a work in progress and all comments are more than welcome.

Startups and Stanford University# Appendix

## The Fields of Activity

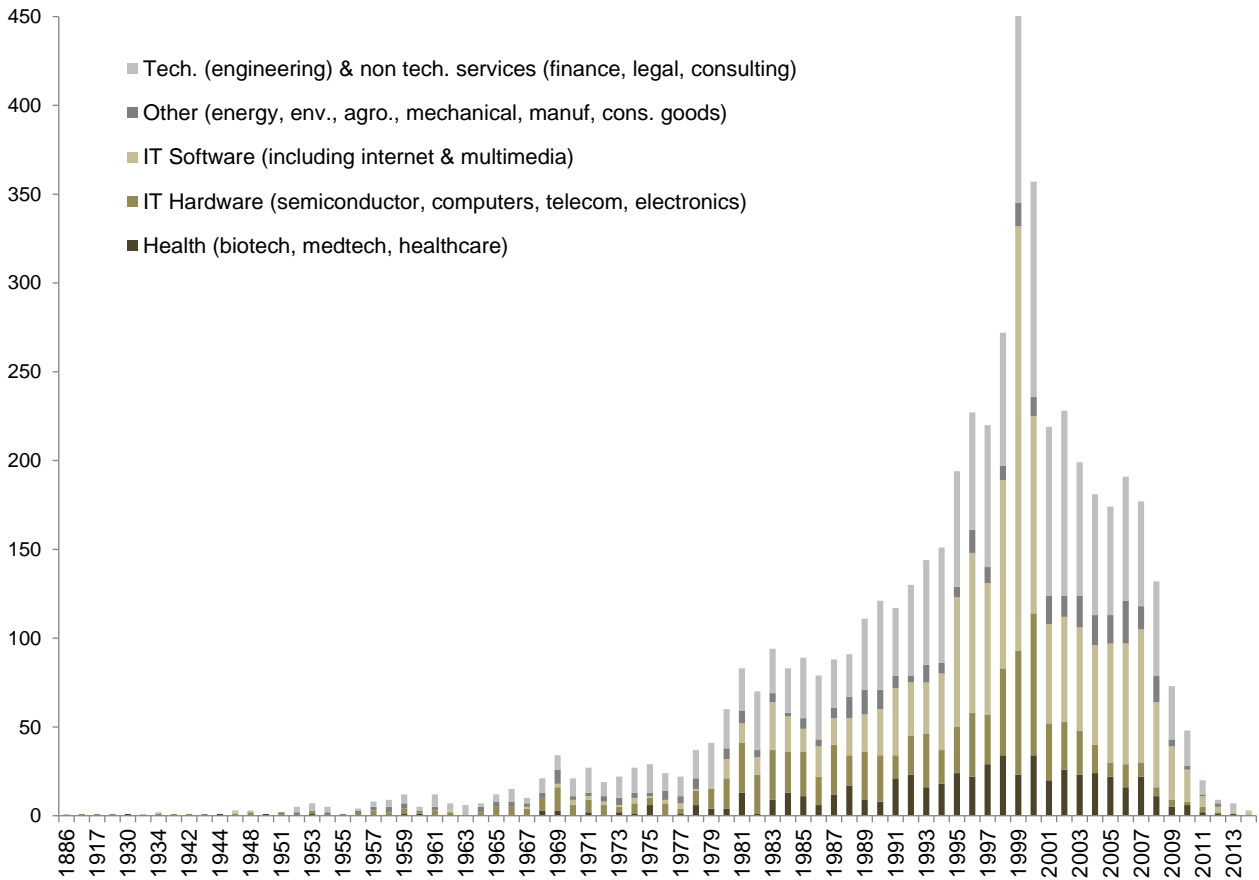

*Figure i: The Stanford startups by year and fields of activity*

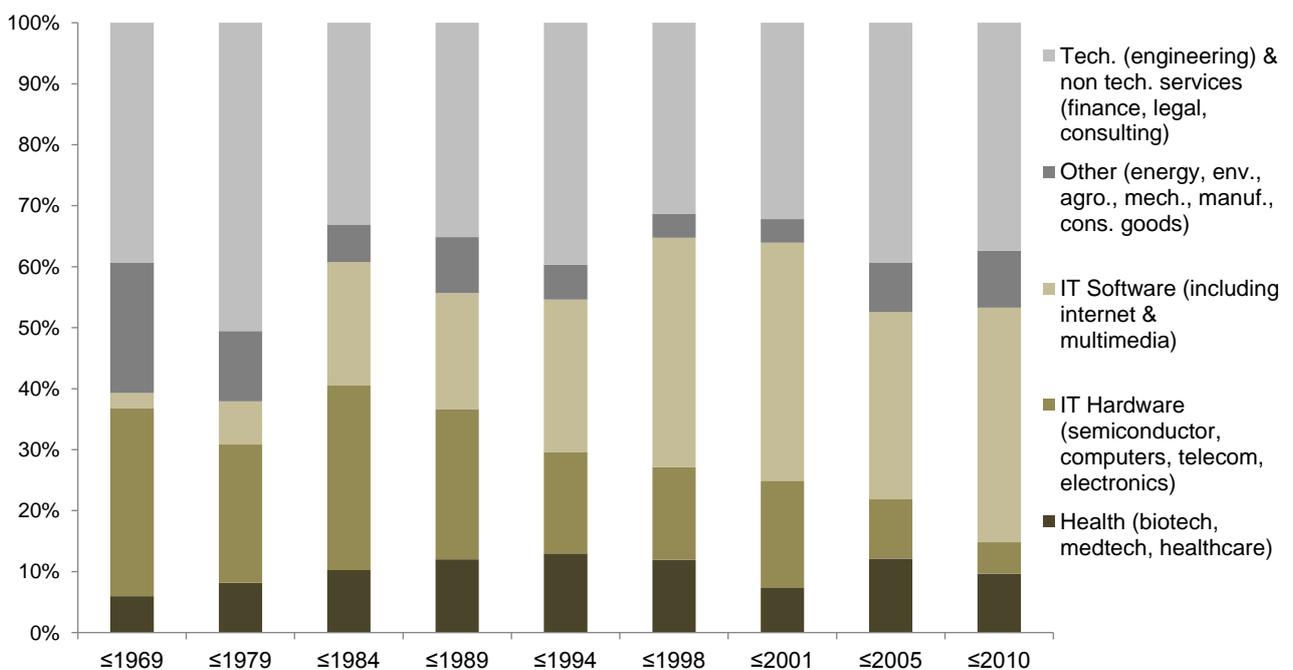

*Figure ii: Ratio of Stanford startups by period of foundation and fields of activity*

July 2017     Page 15

## Life Expectancy and Fields of Activity

| Field of activity | # Firms | Av. years to exit | Field of activity | # Firms | Av. years to exit |
|---|---|---|---|---|---|
| Biotechnology | 258 | 6.6 | Energy | 83 | 10.6 |
| Medtech | 211 | 7.8 | Environment | 36 | 8.8 |
| Healthcare | 98 | 9.6 | Mechanical | 46 | 10.3 |
| Semiconductor | 136 | 7.4 | Manufacturing | 27 | 17.0 |
| Computers | 70 | 7.8 | Multimedia | 109 | 9.8 |
| EDA software | 38 | 6.4 | IT | 175 | 6.9 |
| Telecommunications | 307 | 6.1 | Software | 590 | 6.6 |
| Electronics | 335 | 10.2 | Internet | 745 | 4.0 |
| Optics | 31 | 8.4 | Consumer goods | 184 | 10.0 |
| Entertainment | 115 | 8.2 | Education | 121 | 7.1 |
| Finance | 698 | 9.1 | Engineering | 112 | 14.7 |
| Law | 74 | 10.9 | Tech. services | 114 | 9.3 |
| Non tech. services | 598 | 9.8 | Consulting | 160 | 8.3 |

*Table i: Average time before exit vs. domain of activity*

## Value Creation of Public Companies at IPO

| Fields | ≤1969 | ≤1979 | ≤1984 | ≤1989 | ≤1994 | ≤1998 | ≤2001 | ≤2005 | ≤2010 | Overall |
|---|---|---|---|---|---|---|---|---|---|---|
| Health | 89 | 114 | 111 | 132 | 264 | 419 | 301 | 202 | 237 | 218 |
| IT HW | 127 | 93 | 206 | 214 | 553 | 2193 | 1329 | 3416 | 590 | 611 |
| IT SW |  | 56 | 104 | 767 | 736 | 2165 | 952 | 1793 | 866 | 1276 |
| Other | 70 | 50 | 61 | 1172 | 674 | 191 | 398 | 834 | 1152 | 578 |
| Services | 73 | 272 | 1008 | 2631 | 16 | 159 | 492 | 188 |  | 619 |
| Overall | 109 | 135 | 177 | 419 | 490 | 1743 | 884 | 1020 | 657 | 688 |

*Table ii: Market capitalization at IPO ($M) by fields and periods of foundation*

| Fields | ≤1969 | ≤1979 | ≤1984 | ≤1989 | ≤1994 | ≤1998 | ≤2001 | ≤2005 | ≤2010 | Overall |
|---|---|---|---|---|---|---|---|---|---|---|
| Health | 63 | 135 | 170 | 110 | 196 | 289 | 281 | 272 | 154 | 188 |
| IT HW | 121 | 74 | 225 | 175 | 671 | 1056 | 696 | 3723 |  | 430 |
| IT SW |  | 61 | 213 | 682 | 475 | 2993 | 751 | 1918 | 1160 | 1532 |
| Other | 126 | 129 | 67 | 1707 | 468 | 754 | 382 | 867 | 627 | 725 |
| Services | 384 | 263 | 1056 | 6679 | 13 | 144 | 469 | 350 |  | 1227 |
| Overall | 142 | 133 | 224 | 588 | 398 | 1914 | 632 | 1261 | 716 | 735 |

*Table iii: Market capitalization 12 months after IPO ($M) by fields and periods of foundation*

| Fields | ≤1969 | ≤1979 | ≤1984 | ≤1989 | ≤1994 | ≤1998 | ≤2001 | ≤2005 | ≤2010 | Overall |
|---|---|---|---|---|---|---|---|---|---|---|
| Health | 3 | 8 | 23 | 29 | 33 | 21 | 6 | 10 | 3 | 136 |
| IT HW | 20 | 14 | 37 | 29 | 19 | 21 | 8 | 2 | 1 | 151 |
| IT SW |  | 5 | 14 | 9 | 26 | 51 | 19 | 7 | 4 | 135 |
| Other | 5 | 1 | 2 | 6 | 2 | 1 | 2 | 2 | 1 | 22 |
| Services | 3 | 9 | 2 | 3 | 1 | 1 | 1 | 2 |  | 22 |
| Overall | 31 | 37 | 78 | 76 | 81 | 95 | 36 | 23 | 9 | 466 |

*Table iv: Number of companies taken into account in tables ii and iii*

## M&A Values

| Fields | ≤1969 | ≤1979 | ≤1984 | ≤1989 | ≤1994 | ≤1998 | ≤2001 | ≤2005 | ≤2010 | Overall |
|---|---|---|---|---|---|---|---|---|---|---|
| Health | 360 | 212 | 684 | 105 | 103 | 68 | 94 | 186 | 106 | 142 |
| IT HW | 508 | 165 | 56 | 120 | 78 | 327 | 207 | 152 | 617 | 217 |
| IT SW |  | 415 | 115 | 133 | 150 | 133 | 121 | 172 | 213 | 148 |
| Other | 132 |  | 255 | 119 | 6 | 106 |  | 18 |  | 127 |
| Services | 122 | 719 | 78 | 39 | 156 | 178 | 67 |  |  | 162 |
| Overall | 356 | 265 | 144 | 116 | 113 | 182 | 158 | 167 | 234 | 173 |

*Table v: Average M&A Values ($M) by fields and periods of foundation*

| Fields | ≤1969 | ≤1979 | ≤1984 | ≤1989 | ≤1994 | ≤1998 | ≤2001 | ≤2005 | ≤2010 | Overall |
|---|---|---|---|---|---|---|---|---|---|---|
| Health | 1 | 4 | 4 | 9 | 16 | 24 | 8 | 14 | 5 | 85 |
| IT HW | 12 | 10 | 20 | 18 | 22 | 48 | 46 | 16 | 3 | 195 |
| IT SW |  | 2 | 10 | 11 | 23 | 83 | 38 | 27 | 24 | 218 |
| Other | 4 |  | 2 | 3 | 1 | 1 |  | 1 |  | 12 |
| Services | 4 | 2 | 5 | 2 | 5 | 2 | 3 |  |  | 23 |
| Overall | 21 | 18 | 41 | 43 | 67 | 158 | 95 | 58 | 32 | 533 |

*Table vi: Number of companies for which M&A value is known*

## M&A Values of Companies which had been Publicly Quoted

| Fields | ≤1969 | ≤1979 | ≤1984 | ≤1989 | ≤1994 | ≤1998 | ≤2001 | ≤2005 | ≤2010 | Overall |
|---|---|---|---|---|---|---|---|---|---|---|
| Health | 3.9 | 0.7 | 1.9 | 0.3 | 1.9 | 0.5 | 2.7 | 0.5 |  | 1.3 |
| IT HW | 2.6 | 1.1 | 1.8 | 0.5 | 0.4 | 0.8 | 3.1 | 3.5 |  | 1.3 |
| IT SW |  | 0.5 | 0.5 | 2.0 | 1.5 | 1.9 | 0.6 | 9.9 | 2.1 | 1.7 |
| Other | 1.0 |  | 0.2 | 4.6 |  | 0.1 | 0.5 |  |  | 1.8 |
| Services |  | 0.4 |  |  | 0.7 |  |  |  |  | 0.4 |
| Overall | 2.7 | 0.8 | 1.5 | 0.9 | 1.5 | 1.3 | 1.4 | 5.0 | 2.1 | 1.5 |

*Table vii: Average M&A Values ($B) by fields and periods of foundation*

| Fields | ≤1969 | ≤1979 | ≤1984 | ≤1989 | ≤1994 | ≤1998 | ≤2001 | ≤2005 | ≤2010 | Overall |
|---|---|---|---|---|---|---|---|---|---|---|
| Health | 3 | 5 | 18 | 19 | 19 | 10 | 2 | 3 |  | 79 |
| IT HW | 10 | 9 | 24 | 17 | 6 | 14 | 3 | 1 |  | 84 |
| IT SW |  | 5 | 9 | 8 | 19 | 25 | 8 | 3 | 2 | 79 |
| Other | 1 |  | 1 | 2 |  | 1 | 1 |  |  | 6 |
| Services |  | 2 | 1 |  | 1 |  |  |  |  | 4 |
| Overall | 14 | 21 | 53 | 46 | 45 | 50 | 14 | 7 | 2 | 252 |

*Table viii: Number of companies for which M&A value is known*

## Venture Capital

| Fields | ≤1969 | ≤1979 | ≤1984 | ≤1989 | ≤1994 | ≤1998 | ≤2001 | ≤2005 | ≤2010 | Other | Overall |
|---|---|---|---|---|---|---|---|---|---|---|---|
| Health |  | 6 | 16 | 20 | 35 | 52 | 52 | 60 | 33 | 33 | 40 |
| IT HW | 12 | 12 | 25 | 24 | 37 | 54 | 66 | 65 | 68 | 11 | 47 |
| IT SW |  | 9 | 12 | 30 | 20 | 39 | 35 | 55 | 47 | 59 | 39 |
| Other | 12 | 9 |  | 34 | 28 | 13 | 104 | 174 | 57 |  | 73 |
| Services |  | 20 | 21 |  | 1 | 24 | 12 | 21 | 43 |  | 19 |
| Overall | 12 | 10 | 21 | 24 | 30 | 45 | 47 | 62 | 48 | 40 | 42 |

*Table ix: Average amount of venture capital raised by field and period of foundation*

| Fields | ≤1969 | ≤1979 | ≤1984 | ≤1989 | ≤1994 | ≤1998 | ≤2001 | ≤2005 | ≤2010 | Other | Overall |
|---|---|---|---|---|---|---|---|---|---|---|---|
| Health |  | 9 | 25 | 43 | 60 | 61 | 40 | 61 | 31 | 2 | 332 |
| IT HW | 7 | 13 | 66 | 55 | 57 | 98 | 128 | 41 | 17 | 2 | 484 |
| IT SW |  | 3 | 20 | 23 | 64 | 177 | 213 | 117 | 107 | 2 | 726 |
| Other | 1 | 1 | 1 | 5 | 4 | 4 | 8 | 9 | 12 |  | 45 |
| Services | 1 | 3 | 2 | 0 | 4 | 3 | 10 | 2 | 2 |  | 27 |
| Overall | 9 | 29 | 114 | 126 | 189 | 343 | 399 | 230 | 169 | 6 | 1'614 |

*Table x: Number of companies which raised venture capital*

## Serial Entrepreneurs

| Serial index | Amount raised | M&A Value | Value at IPO | Value 12 m. after IPO |
|---|---|---|---|---|
| 0 | 29 | 159 | 146 | 172 |
| 1 | 8 | 151 | 56 | 55 |
| 2 | 17 | 111 | 77 | 61 |
| 3 | 6 | 18 | 21 | 37 |
| 4 | 3 | 6 | 10 | 9 |
| 2+ | 29 | 148 | 118 | 112 |

*Table xi: Total value creation ($B) and serial entrepreneurs*

| Serial index | Amount raised | M&A Value | Value at IPO | Value 12 m. after IPO |
|---|---|---|---|---|
| 0 | 699 | 331 | 212 | 208 |
| 1 | 293 | 204 | 115 | 115 |
| 2 | 358 | 160 | 91 | 90 |
| 3 | 113 | 44 | 23 | 23 |
| 4 | 52 | 24 | 12 | 12 |
| 2+ | 569 | 251 | 139 | 138 |

*Table xii: Number of firms counted for value creation and serial entrepreneurs*

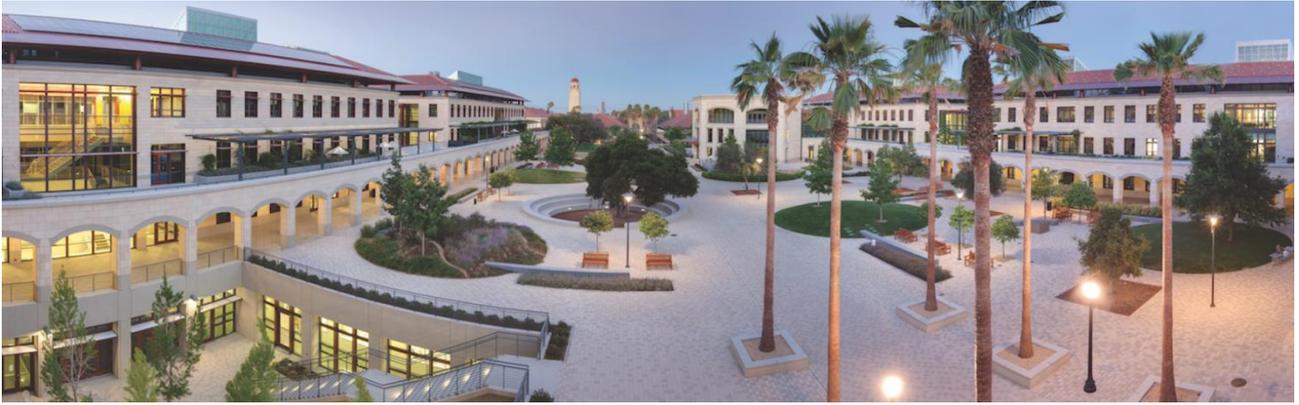

**Startups and Stanford University**

Startups have become in less than 50 years a major component of innovation and economic growth. Silicon Valley has been the place where the startup phenomenon was the most obvious and Stanford University was a major component of that success. Companies such as Google, Yahoo, Sun Microsystems, Cisco, Hewlett Packard had very strong links with Stanford but even these vary famous success stories cannot fully describe the richness and diversity of the Stanford entrepreneurial activity. This report explores the dynamics of more than 5'000 companies founded by Stanford University alumni and staff, through their value creation, their field of activities, their growth patterns and more. The report also explores some features of the founders of these companies such as their academic background or the number of years between their Stanford experience and their company creation.

**About the author**

Hervé Lebret has been working in the startup world for more than 20 years. Since 2005, he has been in charge of support to startup creation at EPFL, the Swiss Federal Institute of Technology in Lausanne. He was before with Index Ventures, a pan-European venture capital firm which invested in Skype, mysql, Numeritech, Virata, Genmab. He used that experience to write in 2007 the book "Start-Up, what we may still learn from Silicon Valley" and the blog www.startup-book.com. Since 2010, he has also been doing research on high-tech startups with a particular focus on Silicon Valley and Stanford University. Lebret was trained in science and engineering, he is a graduate of Ecole Polytechnique (1987) and Stanford University (1990). He did his PhD in 1994 on the topic of convex optimization and its applications, which he still teaches at EPFL in addition to teaching entrepreneurship. He was a researcher in applied mathematics until he switched to venture capital in 1997.